\documentclass[12pt]{article}

\usepackage{cite}
\usepackage{epsfig}

\usepackage{times}

\usepackage{amssymb}
\usepackage{graphicx}

\newcommand{\aver}[1]{\ensuremath{\langle {#1} \rangle}}

\newcommand{\ket}[1]{\left|  #1 \right\rangle}
\newcommand{\tamp}[0]{\ensuremath{t}}
\newcommand{\F}[0]{\ensuremath{\mathcal{F}}}
\newcommand{\re}[0]{\ensuremath{\mathrm{Re}}}
\newcommand{\lraw}[0]{\ensuremath{\leftrightarrow}}
\newcommand{\raw}[0]{\ensuremath{\rightarrow}}
\newcommand{\OD}[0]{\ensuremath{\mathcal{N}}}
\newcommand{\DA}[0]{\ensuremath{\tilde{\Delta}}}
\newcommand{\dc}[0]{\ensuremath{\tilde{\delta}}}
\newcommand{\etaeff}[0]{\ensuremath{\widetilde{\eta}}}

\newenvironment{sciabstract}{%
\begin{quote} \bf}
{\end{quote}}

\newcounter{lastnote}

\title{Vacuum-Induced Transparency}

\author
{Haruka Tanji-Suzuki$^{1,2\ast}$, Wenlan Chen$^{2}$, Renate Landig$^{2}$, Jonathan Simon$^{1}$\\
and Vladan Vuleti\'{c}$^{2}$\\
\normalsize{$^{1}$ Department of Physics, Harvard University, Cambridge, Massachusetts 02138, USA}\\
\normalsize{$^{2}$ Department of Physics,
MIT-Harvard Center for Ultracold Atoms,}\\
\normalsize{and Research Laboratory of Electronics, Massachusetts Institute of Technology,}\\
\normalsize{Cambridge, Massachusetts 02139, USA}\\
\\
\normalsize{$^\ast$To whom correspondence should be addressed; E-mail: tanji@mit.edu.}
}

\date{}

\begin{document}


\baselineskip24pt


\maketitle


\begin{sciabstract}
Photons are excellent information carriers but normally pass through each other without consequence. Engineered interactions between photons would enable applications from quantum information processing to simulation of condensed matter systems. Using an ensemble of cold atoms strongly coupled to an optical cavity, we demonstrate experimentally that the transmission of light through a medium may be controlled with few photons and even by the electromagnetic vacuum field.
The vacuum induces a group delay of 25 ns on the input optical pulse, corresponding to a light velocity of 1600~m/s, and a transparency of 40\% that increases to 80\% when the resonator is filled with 10 photons. This strongly nonlinear effect provides prospects for advanced quantum devices such as photon-number-state filters.
\end{sciabstract}

The experimental realization of strong coherent interactions between individual photons will enable a variety of applications ranging from quantum computing \cite{Pellizzari95,Turchette95,Rauschenbeutel99} to studies of strongly-correlated many-body quantum systems \cite{Chang08}. Two main approaches to generating photon-photon interactions are strong coupling of single emitters to optical resonators \cite{Rempe91,Thompson92,Turchette95,Muenstermann99,Rauschenbeutel99,Birnbaum05,Fushman08} and electromagnetically induced transparency (EIT) in ensembles of atoms \cite{Harris89,Harris98,Fleischhauer05}. Single emitters strongly coupled to resonators can provide substantial optical nonlinearity at the expense of typically large input-output coupling losses and the technical challenges of trapping and manipulating single particles. EIT in atomic ensembles provides an impressive degree of coherent control in simple, elegant experiments\cite{Boller91,Kash99,Hau99,Fleischhauer05}, but the nonlinearities
achieved so far are relatively weak, requiring, e.g., $\sim 500$ photons for all-optical switching\cite{Bajcsy09}. We demonstrate that by using an optical cavity to enhance the EIT control field, the resonant transmission of light through an atomic ensemble can be substantially altered by a few photons and even the cavity vacuum \cite{Field93,Rice96}. As the effect is nonlinear in both control and probe fields at the single-photon level, it should enable advanced quantum optical devices such as photon-number-state filters \cite{Nikoghosyan10} and non-destructive photon-number-resolving detectors \cite{Guerlin07,Schuster07}.
We call the limiting case with no photons initially in the cavity ``vacuum-induced transparency (VIT)'' \cite{Field93} to distinguish it from recent cavity EIT demonstrations using a single atom with cavity-enhanced absorption and a classical control field containing many photons \cite{Muecke10,Kampschulte10}. In contrast, for VIT, the entire system contains at most one photon.

\begin{figure}
\begin{center}
\begin{tabular}{ll}
\begin{minipage}[t]{3.1in}
\begin{center}
\vspace{1pt}
\epsfig{figure=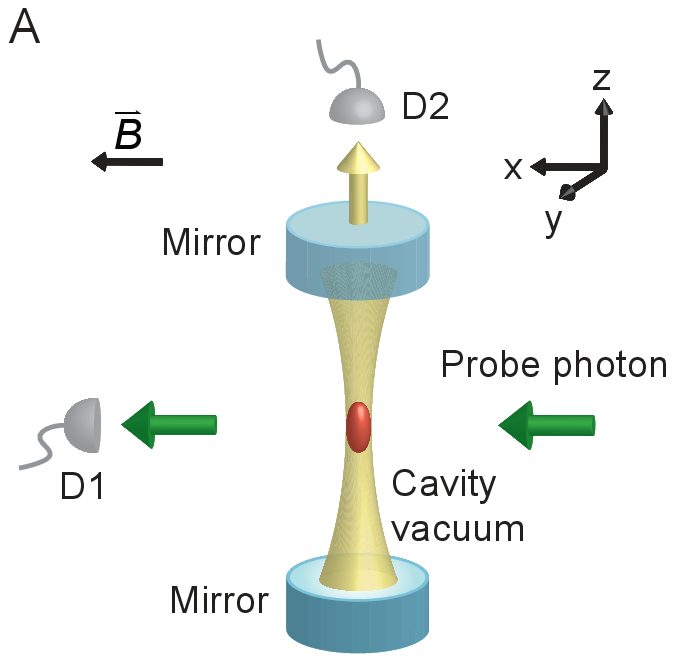,width=3.1in}
\end{center}
\end{minipage}&
\begin{minipage}[t]{2.5in}
\begin{center}
\vspace{0pt}
\epsfig{figure=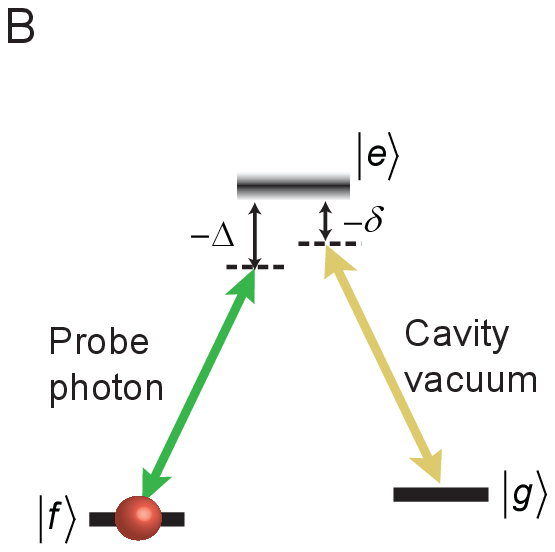,width=2.5in}
\end{center}
\end{minipage}
\end{tabular}
\end{center}
\caption{\label{Fig:Setup} Setup (A) and atomic level scheme (B) for observing vacuum-induced transparency. An ensemble of laser cooled $^{133}$Cs atoms is trapped inside an optical resonator operating in the single-atom strong-coupling regime (cooperativity parameter $\eta>1$). The atoms are prepared in state $\ket{f}$ by optical pumping. The absorption of a probe laser on the transition $\ket{f} \lraw \ket{e}$ is substantially altered when a cavity mode on the transition $\ket{g} \lraw \ket{e}$ is tuned near two-photon resonance. In spite of the cavity mode subtending only a very small ($\sim 10^{-4}$~sr) solid angle along a direction transverse to the probe beam, its vacuum field can substantially reduce the probe absorption by quantum interference. Photon counters D1 and D2 are used to measure the probe transmission and the scattering into the cavity, respectively.}
\end{figure}

We experimentally realize Field's original proposal \cite{Field93} to replace the EIT control field by the vacuum field inside a strongly coupled cavity (Fig.\ \ref{Fig:Setup}). In an atomic $\Lambda$ system $\ket{f} \lraw \ket{e} \lraw \ket{g}$ with two stable states $\ket{f},\ket{g}$, the probe beam addresses the $\ket{f} \raw \ket{e}$ transition, while the cavity mode is tuned near the $\ket{g} \raw \ket{e}$ transition.  A cold atomic ensemble is prepared in the state $\ket{f}$ by optical pumping. VIT for the probe beam can be thought of as arising from a vacuum-induced Raman process where the incoming probe photon is absorbed, quickly emitted into the cavity, then reabsorbed by the ensemble, and re-emitted collectively back into the probe mode. Thus the incoming probe photon creates its own transparency by destructive interference in the excited state $\ket{e}$ arising from the two transitions $\ket{f} \raw \ket{e}$ and $\ket{g} \raw \ket{e}$. In contrast to standard EIT, here the effective control field on the $\ket{g} \raw \ket{e}$ transition depends sensitively on the photon number in the probe field.
When there are several photons in the probe field, those photons are coupled to the cavity mode,
constituting an effective probe-power-dependent control field for the VIT process.
As the EIT group delay depends on the control coupling strength \cite{Fleischhauer05}, different probe Fock states experience different group delay and therefore an incoming coherent-state probe pulse may be resolved into a train of photon-number components \cite{Nikoghosyan10}.

VIT requires strong coupling between a single atom and a cavity, i.e., a single-atom cooperativity parameter $\eta_0=4g^2/(\kappa \Gamma)$ exceeding unity.
Here $2g$, $\kappa$, and $\Gamma$ are the single-photon Rabi frequency, cavity linewidth, and atomic linewidth (FWHM), respectively. For unity oscillator strength the cooperativity parameter is a geometric quantity associated with the cavity characteristics alone, and can be written in terms of the finesse $\F$, waist $w$ and wavenumber $k$ of the cavity mode as $\eta_0=24 \F/(\pi k^2 w^2)$ \cite{Tanji-Suzuki11}. Our parameters $\lambda=2\pi/k = 852$~nm, $w=35 \mathrm{\mu}$m and $\F=6.3(5) \times 10^4$ yield a maximum cooperativity for a single $^{133}$Cs atom at an antinode of $\eta_0=7.2(5)$. The actual cooperativity $\eta$ available in the experiment is smaller, due to oscillator strength $f_{eg}<1$ for the $\ket{g} \lraw \ket{e}$ transition in question, and spatial averaging of the coupling along the standing-wave resonator mode.

For probe light illuminating the ensemble from the side (Fig.\ \ref{Fig:Setup}), the amplitude transfer function $\tamp=e^{i k L \chi/2}$ can be expressed in terms of the susceptibility $\chi$ that in the limit of weak coupling on the probe transition (single probe photon) is given by \cite{Field93,Rice96,Fleischhauer05}
\begin{equation}
\chi=-\frac{\OD}{k L}\frac{\DA-\left( \eta - \DA \dc \right) \dc - i \left(\eta +1 + \dc^2\right)}
{\left(\eta+1 - \DA \dc \right)^2 + \left( \DA+\dc \right)^2}.
\label{Eq:Susceptibility}
\end{equation}
Here, $\OD$ is the resonant optical depth of the ensemble with length $L$ along the probe beam, and $\DA=2\Delta/\Gamma=2(\omega_p-\omega_{ef})/\Gamma$ and $\dc=2(\Delta-\delta)/\kappa=2(\omega_p-\omega_c-\omega_{gf})/\kappa$ are the normalized probe-atom detuning and the ``two-photon" detuning, respectively (Fig.\ \ref{Fig:Setup}B), where $\omega_p,\omega_c,\omega_{ji}$ are the frequencies of the probe, cavity mode, and atomic transition $\ket{i}\raw \ket{j}$, respectively.
Eq. \ref{Eq:Susceptibility} can be obtained from the standard EIT expression \cite{Fleischhauer05} with states $\ket{f;n_p=1;n_c=0}$, $\ket{e;0;0}$, $\ket{g;0;1}$, where $n_p$ and $n_c$ are the probe and cavity photon numbers, respectively, with the cavity linewidth $\kappa$ assigned to the state $\ket{g;0;1}$ \cite{Field93}.
When both the probe field and the cavity mode are resonant with their respective atomic transitions, $\Delta=\delta=0$, the transmission probability is given by $ \left| \tamp \right|^2= e^{-\frac{\OD}{\eta+1}}$, i.e., the resonant optical depth $\OD$ is reduced by a factor $\eta+1$ by the cavity vacuum field.

The observation of VIT requires substantial atomic absorption in a transverse direction for an optical resonator that operates in the strong coupling limit $\eta>1$ for a single atom \cite{Rempe91,Muenstermann99,Birnbaum05}. This parameter regime has recently been achieved with Bose-Einstein condensates in resonators with small mode volume \cite{Gupta07,Brennecke07,Colombe07}. Here we use a relatively long (1.4~cm) cavity that allows us to operate a magneto-optical trap for $^{133}$Cs inside the cavity, and directly load up to $10^5$ atoms into a far-off resonance optical-lattice trap operated at $937$~nm inside the resonator. The three-level system is chosen as $\ket{f} \equiv \ket{6S_{1/2},F=3,m_F=3} $, $\ket{e} \equiv \ket{6P_{3/2},4,4}$, $\ket{g} \equiv \ket{6S_{1/2},4,4}$ to provide a good combination of oscillator strengths in both arms ($f_{ef}=0.42,f_{eg}=0.47$).
The quantization axis is defined by a 1.6~G magnetic field along the propagation direction ($x$) of the probe beam (Fig.\ \ref{Fig:Setup}).
The $\sigma^+$-polarized probe beam is tightly focused  by an aspheric lens to a waist $w_p=2.3\ \mathrm{\mu}$m at the cavity mode. We achieve an optical depth up to $\OD=0.4$ by optically pumping all atoms into the $F=3$ hyperfine manifold, and more than 90\% into state $\ket{f}$. The thickness of the cloud along the probe beam is $L=20\ \mathrm{\mu}$m at a typical temperature of $100\ \mathrm{\mu}$K, with an estimated peak atomic density of $1.1\times10^{11}\ \mathrm{cm}^{-3}$. Typically $\sim 20$ atoms are contained in the volume defined by the probe beam.

\begin{figure}
\centering{\includegraphics[width=6in]{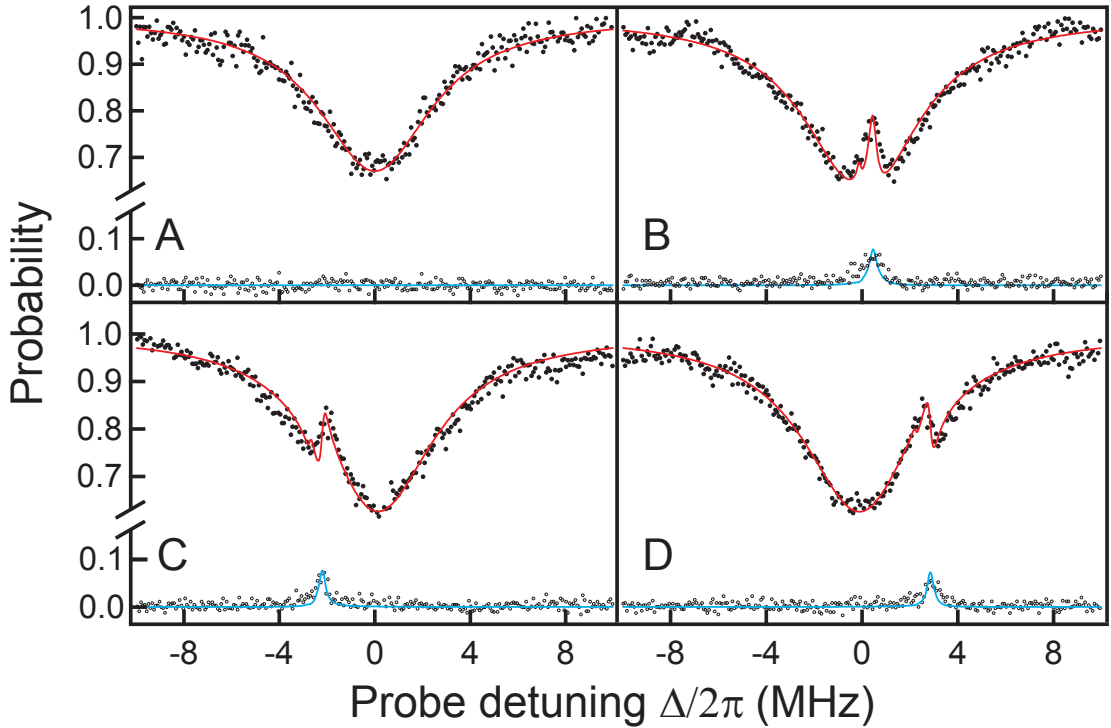}}
\caption{\label{Fig:Spectra} Atomic absorption spectrum (A) and VIT spectra (B-D) for different cavity-atom detunings: (B) $\delta/(2\pi)=0.5$~MHz, (C) $\delta/(2\pi)=-2.2$~MHz, (D) $\delta/(2\pi)=2.8$~MHz. The transmission probability (upper curves) and the probability of emission into the cavity (lower curves) are measured simultaneously versus probe-atom detuning $\Delta$ by photon-counting detectors D1, D2. Near the two-photon resonance $\Delta \approx \delta$ the absorption is suppressed by VIT, and a fraction of the incoming photons is redirected into the cavity. Data for both processes for all values of $\delta$ are simultaneously described by the VIT model described in the text (solid lines).}
\end{figure}

With the cavity mode tuned far off resonance from the $\ket{g} \lraw \ket{e}$ transition, the probe frequency is scanned across the $ \ket{f} \raw \ket{e}$ resonance, revealing a Lorentzian absorption profile with a linewidth of 5.46(7)~MHz (Fig.\ \ref{Fig:Spectra}A), where the slight broadening over the natural linewidth $\Gamma/(2\pi)= 5.2$~MHz is due to the laser linewidth. When the cavity mode is tuned close to the $\ket{g} \lraw \ket{e}$ transition, a transparency window opens up around the $ \ket{f} \lraw \ket{g}$ two-photon transition frequency (Figs.\ \ref{Fig:Spectra}B-D). To prevent the accumulation of atoms incoherently pumped by the trapping light into the $F=4$ hyperfine state whose absorption would spoil the cavity finesse, the probe field is turned on and off every $4\ \mathrm{\mu}$s during the 2.5 ms long frequency scan, and a depumping beam emptying the $F=4$ hyperfine state is turned on during the probe dark times. The 4-$\mathrm{\mu}$s duration of the probe pulses is chosen so that the modulation-induced frequency broadening is smaller than the cavity linewidth $\kappa/(2\pi) = 173(13)$~kHz.
To probe the steady-state response of the system as described by Eq. \ref{Eq:Susceptibility} , we restrict the analysis to times $t \geq 0.5\ \mathrm{\mu}$s where transients associated with the width of the transparency window $(1+\eta)\kappa$ have decayed.
At the probe power of $220\ \mathrm{fW}$ and the optical depth of $\OD=0.4$, we post-select data for $t< 2.6\ \mathrm{\mu}$s such that the total number of absorbed photons is $0.8<1$.

The VIT spectra for various atom-cavity detunings, as well as the accompanying photon leakage from the cavity mode, as shown in Figs.\ \ref{Fig:Spectra}B-D, can be simultaneously fit to the VIT model, Eq.\ \ref{Eq:Susceptibility}.
While the vacuum Rabi splitting is not observable in our parameter regime ($\eta > 1$ but $2 g<\Gamma$), the transparency in a narrow window is clearly enhanced by quantum interference.
The observed resonance is slightly broader than predicted by the model, due to a small, independently observed line broadening of 200~kHz that is caused by atom-induced shifts of the cavity mode frequency that fluctuate with the number of loaded atoms. The spectra also reveal a small contribution from the four times weaker VIT transition $\ket{f} \lraw \ket{e} \leftrightarrow \ket{6S_{1/2},4,3}$ that is two-photon Zeeman shifted by 0.6~MHz.

\begin{figure}
\centering{\includegraphics[width=5in]{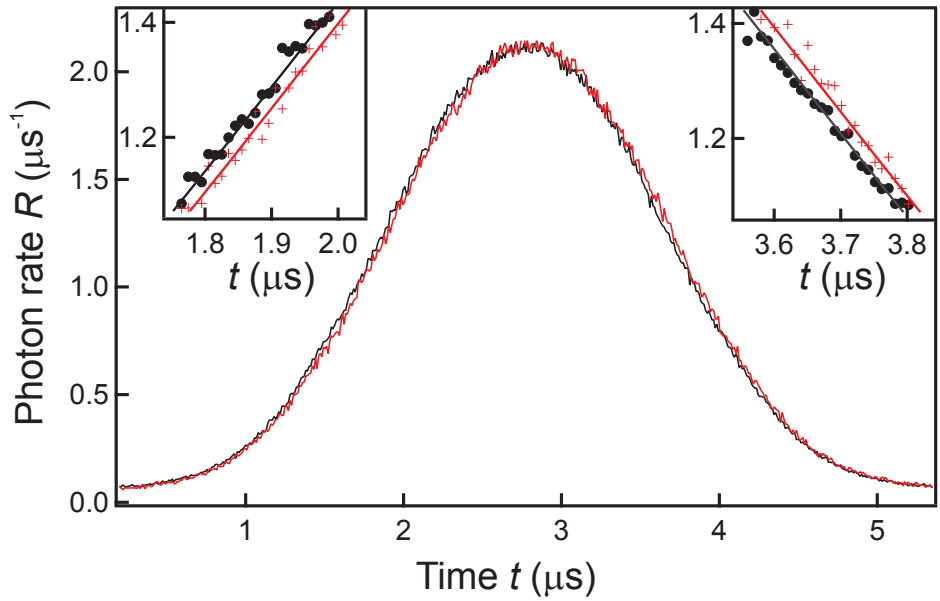}}
\caption{\label{Fig:GroupDelay} Vacuum-induced group delay of a probe pulse. The black circles show the probe pulse in the absence of atoms, the red crosses indicate the probe pulse traveling through the atomic medium on VIT resonance $\Delta=\delta=0$. The observed delay induced by the cavity vacuum field is $\tau=25(2)$~ns. The delayed pulse experiences absorption and has been rescaled by a factor 1.6 for easier visualization of the group delay.}
\end{figure}

As in standard EIT, the index of refraction $n=\sqrt{1+\re(\chi)}$ is unity on resonance $\Delta=\delta=0$, and varies sharply with probe frequency $\omega_p$ for fixed cavity detuning $\delta=0$, giving rise to a reduced probe group velocity $v_g=c/\left( n+\omega_p\frac{\mathrm d n}{\mathrm d\omega_p} \right)  \ll c$ \cite{Harris98,Hau99,Fleischhauer05}. Pulses that are sufficiently narrow spectrally to fit into the transparency window should therefore  according to Eq.\ \ref{Eq:Susceptibility} experience a maximum group delay \cite{Fleischhauer05} of $\tau_\mathrm{max}=\frac{\OD}{\kappa}\frac{\eta}{(\eta+1)^2}$.
During this delay time, the incoming probe photon is in part stored as a stationary photon inside the optical resonator for up to a time $\kappa^{-1}$, while a spin excitation with one atom in state $\ket{g}$ is simultaneously created in the ensemble.
The delay decreases with increasing $\eta$, because a stronger control field reduces the population in state $\ket{g}$, and a smaller fraction of the photon is stored in the cavity accordingly.
Figure \ref{Fig:GroupDelay} shows a Gaussian pulse of $T_P=1.73\ \mathrm{\mu}$s duration that is delayed by the vacuum by $\tau=25(2)$~ns, close to the value $35$~ns calculated from Eq.\ \ref{Eq:Susceptibility} for the measured optical depth $\OD=0.5$ (achieved in a double-pass geometry). The small discrepancy is explained by small ($\sim 200$~kHz) atom-induced fluctuations of the cavity resonance frequency as described above. While the absolute delay $\tau$, corresponding to a group velocity of $v=1600$~m/s, is small in the present system due to the relatively small optical depth $\OD$, the observation nevertheless establishes experimentally that a vacuum input control field can delay a probe pulse. Larger delays can be achieved by increasing $\OD$, either by enhancing the atomic density via further optical cooling \cite{Vuletic98}, or by means of a multi-pass geometry for the probe beam.

\begin{figure}
\centering{\includegraphics[width=5in]{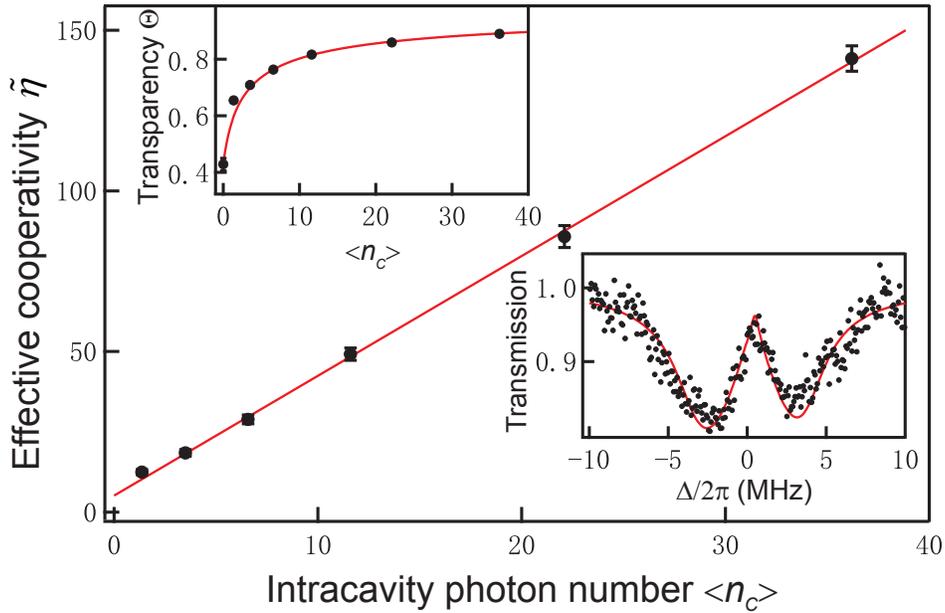}}
\caption{\label{Fig:ControlNL} VIT as a limiting case of EIT: Effective cooperativity $\etaeff$ at an antinode as a function of average cavity photon number $\aver{n_c}$ from fits to measured spectra as shown in the right lower inset for $\aver{n_c}=22$. The effective cooperativity is expected to scale as $\etaeff=\etaeff_0(\aver{n_c}+1)$, and we find good agreement with a linear fit. The upper left inset shows the peak transparency $\Theta$ vs. $\aver{n_c}$, demonstrating that even one control photon substantially changes the transmission. Error bars indicate $\pm 1\sigma$ standard deviation.}
\end{figure}

Unlike standard EIT, the VIT process is intrinsically nonlinear at the single-photon level: In EIT the classical control field with very large photon number $\aver{n_c} \gg 1$ alone determines the transparency window and group velocity of the probe light \cite{Hau99,Fleischhauer05}; there is no dependence on the weak probe field with photon number $\aver{n_p} \ll \aver{n_c}$.
On the other hand, in a VIT system the control field is initially the vacuum, and the transparency window and group delay vary strongly with $n_p$, which sets $n_c$ as described earlier.
To demonstrate the strong optical nonlinearity intrinsic to the VIT system, we directly vary the average cavity photon number $\aver{n_c}$ by exciting the cavity mode with a weak laser beam, and measure the probe transmission. With the cooperativity $\eta$ replaced by a free parameter, we fit the measured spectra (see lower inset to Fig.\ \ref{Fig:ControlNL} as an example) using Eq.\ \ref{Eq:Susceptibility} and taking into account the spatial variation of the cavity coupling. Fig.\ \ref{Fig:ControlNL} shows the thus extracted effective cooperativity at an antinode $\etaeff$ vs.\ $\aver{n_c}$. Because the control Rabi frequency is given by $\Omega_c=2g\sqrt{n_c+1}$, we expect a linear dependence of $\etaeff$ on $\aver{n_c}$ with a slope $m$ equal to the y-axis intercept $\etaeff_0$. A linear fit to the data for $\aver{n_c}>2$, where the atom-induced cavity line broadening has negligible effect, yields $m=3.7(1), \etaeff_0=5(1)$ and  the ratio $\etaeff_0/m=1.4(3)$, in reasonable agreement with the model that predicts $m=\etaeff_0=f_{eg}\eta_0=3.4$. The upper inset shows the peak transparency $\Theta$ vs.\ $\aver{n_c}$.
The transparency is defined as $\Theta=(T'-T)/(1-T)$, where $T' (T)$ denotes the resonant transmission with (without) the control field, and $T=e^{-\OD}=0.67$.
This plot shows that a substantial transparency increase over the vacuum-control level occurs already for one intracavity photon. In the future, it should be possible to use this effect, e.g., for a non-destructive measurement of the the intracavity photon number \cite{Schuster07,Guerlin07}.

We have demonstrated that a vacuum field can generate a transparency window in an ensemble of three-level atoms, and observed the associated group delay.
By using a cavity-enhanced control field, we could substantially modify the transmission of an atomic ensemble with $\sim10$ control photons.
We also note that two probe beams, even when passing through spatially separated regions of the atomic ensemble, should influence each other's group velocity through the common interaction with the cavity mode, paving the way to cavity-mediated strong photon-photon interaction and quantum gates \cite{Pellizzari95,Turchette95}.
In such a geometry, the technical roadblocks associated with both cavity-coupling losses \cite{Turchette95,Thompson92,Muenstermann99,Fushman08} and motional and state control of single atoms \cite{Nussmann05a,Birnbaum05} are bypassed. More generally, this work offers the prospects of strongly nonlinear, multimode quantum optics, with a realistic outlook for advanced quantum devices operating coherently with single photons.




 This work was supported by the NSF, the NSF funded Center for Ultracold Atoms, DARPA, and the ARO.

\end{document}